\documentclass[runningheads]{llncs}
\usepackage[T1]{fontenc}
\usepackage{url}
\usepackage{caption}
\usepackage{wrapfig}
\usepackage{enumitem}

\usepackage{url}
\usepackage{hyperref}
\usepackage{graphicx} 
\usepackage[most]{tcolorbox}
\usepackage{multirow}
\newtcolorbox{promptbox}{
    colback=gray!5,     
    colframe=black!75,  
    left=1em,          
    right=0.5em,       
    top=0.5em,         
    bottom=0.5em,      
    sharp corners,     
    boxrule=1pt         
}
\usepackage{graphicx}
\begin{document}
\titlerunning{Fediverse Sharing: Cross-platform Interaction Dynamics }

\title{Fediverse Sharing: Cross-Platform Interaction Dynamics between Threads and Mastodon Users}
\author{Ujun Jeong \and Alimohammad Beigi \and Anique Tahir\thanks{Work done prior to joining Amazon} \and \\ Susan Xu Tang \and H. Russell Bernard \and Huan Liu}

\authorrunning{Jeong et al.} 

\tocauthor{Ujun Jeong, Alimohammad Beigi, Anique Tahir, Susan Xu Tang, H. Russell Bernard, Huan Liu}

\institute{Arizona State University, AZ, 85281, USA \\
\email{\{ujeong1, abeigi, artahir, Susan.Tang, asuruss, huanliu\}@asu.edu}}






\maketitle              
\begin{abstract}
Traditional social media platforms, once envisioned as digital town squares, now face growing criticism over corporate control, content moderation, and privacy concerns. Events such as Twitter’s acquisition (now X) and major policy changes have pushed users toward alternative platforms like Mastodon and Threads. However, this diversification has led to user dispersion and fragmented discussions across the walled gardens of social media platforms. To address these issues, federation protocols like ActivityPub have been adopted, with Mastodon leading efforts to build decentralized yet interconnected networks. In March 2024, Threads joined this federation by introducing its Fediverse Sharing service, which enables interactions such as posts, replies, and likes between Threads and Mastodon users as if on a unified platform. Building on this development, we study the interactions between 20,000+ Threads users and 20,000+ Mastodon users over a ten-month period. Our work lays the foundation for research on cross-platform interactions and federation-driven platform integration.



\end{abstract}

\section{Introduction}

Growing frustrations with traditional social media platforms --- rooted in concerns over content moderation, privacy measures, and corporate influence --- have led many users to seek alternative spaces for online interaction~\cite{fiesler2020moving}. This discontent became particularly evident following Twitter’s acquisition on October 27, 2022, which triggered a mass exodus of users exploring alternatives~\cite{jeong2024exploring,cava2023drivers,he2023flocking}. As a result, the social media landscape has become increasingly divided between centralized and decentralized social networking platforms.

Several prominent alternatives have emerged~\cite{jeong2025navigating,jeong2024user}. Threads, a centralized platform by Meta, uses corporate control and algorithmic feeds, with over 130 million users. In contrast, Mastodon is a decentralized, open-source network offering user control over moderation and their data, with over 10 million users. However, this emergence of diverse platforms has led to a more fragmented social media landscape, as user bases become spread across various networks~\cite{di2024characterizing,oxford2024social}.

\begin{figure}[t]
  \centering
  \begin{minipage}[t]{0.48\textwidth}
    \centering
    \includegraphics[width=\textwidth]{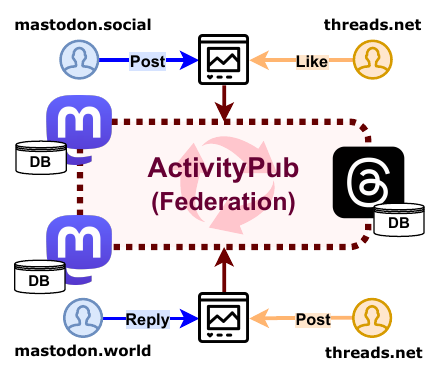}
    \captionsetup{singlelinecheck=off}
    \caption{Federation for cross-platform interactions enabled by the ActivityPub protocol. On the left, Mastodon users from two servers; on the right, Threads users can post, reply, and like. These actions are synced across platforms by replicating data across their databases. In other cases, when multiple users from each platform like the same post, their actions are also shared across Threads and Mastodon.}
    \label{fig:overview_anime}
  \end{minipage}%
  \hfill
  \begin{minipage}[t]{0.48\textwidth}
    \centering
    \includegraphics[width=\textwidth]{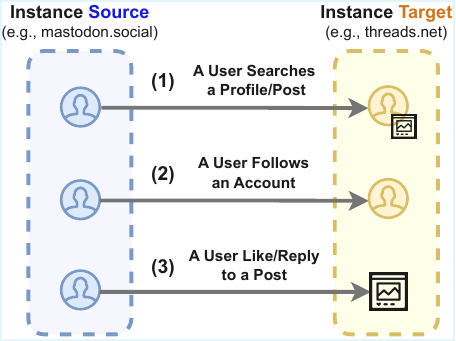}
    \caption{Three key scenarios illustrating how federation is established between ActivityPub-based instances. Federation occurs when a user in a source instance becomes aware of a target instance through one of the following actions: (1) direct search, (2) following a user, or (3) interacting with a post (e.g., liking or replying). Each of these scenarios independently triggers message exchanges between instances.}
    \label{fig:Mastodon_Federation}
  \end{minipage}
\end{figure}


Fortunately, federation-based protocols like ActivityPub have emerged as a solution to the growing fragmentation of social media platforms~\cite{zignani2018follow}. By connecting independent networks through standardized mechanisms, ActivityPub allows users to interact across different instances. Mastodon is an early adopter of ActivityPub, using it to synchronize posts, replies, and likes within its federated network. A major milestone was later reached on March 17, 2024, when Threads adopted ActivityPub through its Fediverse Sharing service, enabling direct interaction between Threads and Mastodon users. Figure~\ref{fig:overview_anime} illustrates how ActivityPub facilitates these cross-platform interactions by synchronizing actions and replicating data across respective platform databases, and Figure~\ref{fig:Mastodon_Federation} depicts how federation is initially established across different instances.

In this paper, we study Fediverse Sharing, the first large-scale cross-platform interactions between Threads and Mastodon users. Collected through Mastodon's public API, the study includes profiles and behaviors of 20,000+ Threads users who opted into FediverseSharing and 20,000+ Mastodon users who interacted with them. Spanning ten months post-launch, it documents posts, replies, and likes shared across the platforms. By examining these interactions, we provide insights into how federation-driven platform integration impacts user engagement, content creation, and intergroup dynamics within a federated network.


Our key contributions are as follows:

\begin{itemize}
\item \textbf{First Cross-platform Interaction Study}: We propose a unique way to collect data for user interactions between Threads and Mastodon, including posts, replies, and likes.

\item \textbf{Impact of Federation-driven Platform Integration}: Our longitudinal data provides an analysis of users' behavioral shifts before and after the introduction of Fediverse Sharing.

\item \textbf{Insights into Federated Social Network}: We analyze different instances within a federated network, highlighting key factors associated with federation and user interactions.
\end{itemize}

\section{Related Work}

\subsection{Platform Integration}
Platform integration is a strategic business move often framed within two-sided markets, where platforms mediate between user groups with complementary interests~\cite{schreieck2024typology,evans2016matchmakers,rochet2003platform}. Such integration enhances accessibility, fosters interaction between distinct platforms, and opens new revenue streams. A notable example is Google’s integration of YouTube, which improves content recommendations, deepens creator-audience engagement, and boosts advertising revenue~\cite{parker2017platform,huang2017social}. Nevertheless, social media platforms have historically emphasized content visibility over direct cross-platform engagement. Facebook’s early integrations with TripAdvisor and Yelp enabled content sharing through Facebook but did not support interaction between users in those services~\cite{cao2024consequences,huang2017social}. Similarly, Facebook and Instagram facilitate content distribution while maintaining distinct user bases, leveraging network effects to increase reach~\cite{abdelkafi2019multi,jeong2021fbadtracker}. Recently, open-source platforms like Mastodon and Bluesky have promoted federation-driven social media, shifting control away from corporate-owned platforms~\cite{la2021understanding,jeong2024bluetempnet}. Protocols like ActivityPub and AT Protocol foster interoperability, allowing different platforms to communicate. However, these integrations raise new challenges regarding governance, content moderation, and the sustainability of platform integration~\cite{zhang2024emergence}.

\subsection{Community Interaction}
Community dynamics shape human behavior by influencing how individuals relate to their group, view outsiders, and respond to changes in community stability~\cite{schmidt2025concept}. Offline research, such as studies on ethnogenesis, explores how interactions between groups contribute to the emergence of new collective identities~\cite{okamura1981situational}. In moments of cultural tension, people actively signal their identity to build trust within their group and maintain boundaries with others~\cite{banerjee2022situational,negron2014new}. Similarly, online communities function as digital tribes, where trust is built through the exchange and reinforcement of community norms~\cite{pettigrew2013does,demoulin2013intergroup}. Platforms like Reddit illustrate how engagement in toxic or partisan subreddits can spread hostility, influencing the behavior of adjacent communities~\cite{russo2024stranger,kumar2018community}. Research on political polarization suggests that exposure to diverse perspectives, including out-group perspectives, can reduce animosity and affective polarization, fostering better understanding of opposing groups~\cite{cohen2010competitive}. While such exposure may lower outgroup animosity, it is often shown to be insufficient to alter polarization. Federated social networks introduce a decentralized model of community interaction, where moderation is handled locally rather than through uniform corporate policies~\cite{schreieck2024typology}. This community-driven approach gives users more autonomy but also leads to inconsistent content regulation and varying levels of trust. While it can empower users, it also raises concerns about platform stability, the presence of harmful content, and ideological conflicts in the absence of centralized enforcement.

\vspace{0.25cm}
We distinguish our study as the first to explore cross-platform interactions in a federated social media ecosystem. Our study reveals how platform integration influences user behavior, particularly during the early stages of integration between structurally distinct platforms. By highlighting how these connections reshape social interactions, we offer new perspectives on a landscape increasingly challenged by user fragmentation, limited access, and diverse governance.

\section{Preliminaries}

\subsection{Federated Social Networking}

Federated social networks operate through a decentralized architecture, in which independently hosted servers (also known as instances) communicate using shared protocols. The cornerstone of this interoperability is ActivityPub, a protocol published by the W3C. It enables people to share posts, replies, likes, and follows across platform boundaries. Each user is modeled as an actor capable of sending and receiving messages, facilitating seamless interactions between compatible services. The Fediverse refers to the federated social networking ecosystem, which includes platforms like Mastodon, PeerTube, and others. Unlike centralized platforms controlled by a single company, the Fediverse empowers individual communities to define their own moderation policies, data ownership models, and user norms. As shown in Figure~\ref{fig:Mastodon_Federation}, federation begins when an instance becomes aware of another through certain user-driven actions. Administrators retain control over federation boundaries by selectively blocking domains or users. 

\subsection{Fediverse Sharing}

Among ActivityPub-compliant platforms, Mastodon is the most widely used, accounting for approximately $58.5\%$ of all instances\footnote{This statistics is based on the Fediverse network tracker at \url{https://fedidb.com/}}. Mastodon is a decentralized microblogging service where users across different servers can interact through federation. In contrast, Threads is a centralized platform developed by Meta and launched on July 5, 2023. Threads entered the Fediverse on March 17, 2024, by enabling a service called Fediverse Sharing. This service allows Threads users to interact with the users in Mastodon in three primary forms as follows:

\begin{itemize}
    \item \textbf{Status (Posts and Replies):} Users can create and share posts and replies across platforms. Mastodon users can also respond to Threads users’ posts.
    \item \textbf{Likes:} Both Mastodon and Threads users can like each other's content, with notifications shared across the two platforms.
    \item \textbf{Follows:} Users on both platforms can follow each other, expanding visibility and reach across network boundaries.
\end{itemize}

Although Threads has improved its interoperability with Mastodon, several limitations remain. Threads users cannot reply to statuses from Mastodon users, which may prevent sequential conversations. Furthermore, due to privacy constraints, Mastodon users cannot view the social graph of Threads accounts.

\section{Data Collection}  

\begin{figure}
\centering
  \includegraphics[width=1.0\textwidth]{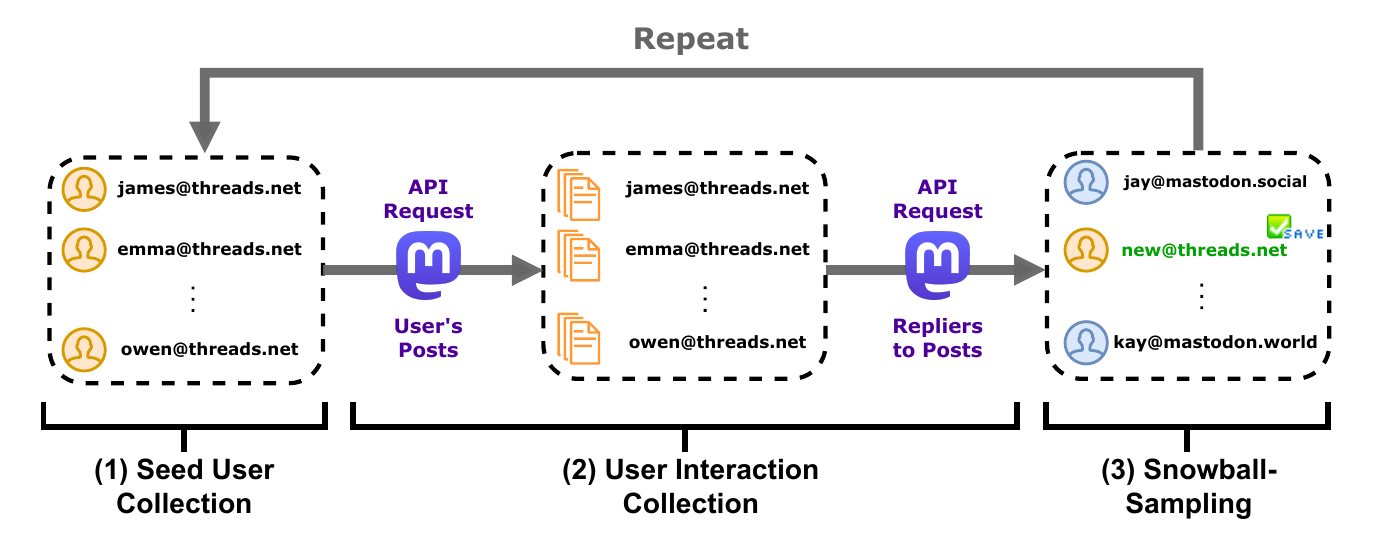}
\caption{Interaction-Driven Snowball Sampling (IDSS) collects Threads users by iteratively tracing repliers with \texttt{@threads.net} domain to the posts of seed users.}
\label{fig:fediversesharing_data_collection}
\end{figure}

The official Threads API\footnote{\url{https://developers.facebook.com/docs/threads/}} does not currently support profile searches or any read permissions. This limitation restricts the collection of users directly on Threads. To address this, we leveraged Mastodon's official API\footnote{\url{https://docs.joinmastodon.org/api/}}, which provides access to Threads users who voluntarily enabled Fediverse Sharing, which makes their profiles and status publicly accessible on Mastodon\footnote{Profile search differs from Threads' keyword search, which requires user approval in Meta Developer App and returns keyword-matching posts. As of July 14, 2025, profile discovery is allowed but limited to public users aged 18+ with 1,000+ followers.}.
Mastodon's API restricts profile searches to $10,000$ accounts, hindering scalable data collection. To address this limitation, we propose \textbf{Interaction-Driven Snowball Sampling (IDSS)}, which consists of three key data collection stages as shown in Figure~\ref{fig:fediversesharing_data_collection}.

We selected \texttt{mastodon.social} as the primary query instance of the Mastodon API due to its extensive federation with other instances. This broad connectivity provides access to a diverse range of users across the Fediverse, including both Threads users and users from other instances federated with \texttt{mastodon.social}.

\begin{enumerate}
    \item \textbf{Seed User Collection:} Seed user collection can be tailored to a study’s needs, including manual sampling from Threads. In our approach, we automated this step by utilizing the \texttt{/api/v2/search} endpoint to find profiles containing \texttt{@threads.net}. After removing samples where the keyword appeared outside the account handle, we identified 9,635 publicly accessible Threads user profiles as our initial seed set.
    
    \item \textbf{User Interaction Collection:} We retrieved the complete set of posts from Threads users through the \texttt{/api/v1/accounts/:id/statuses} endpoint, along with all replies associated with those posts. Note that Threads users' posts are accessible on Mastodon only after they enable Fediverse Sharing, and only posts made afterward are available.

    \item \textbf{Stage 3: Snowball Sampling through User Interaction:} As Mastodon's API does not expose the social graph of Threads users, we adopted an interaction-driven approach. We identified additional Threads users by filtering accounts ending in \texttt{@threads.net} that had replied to posts from seed users. These users were then added to the seed set for iterative expansion.
\end{enumerate}


As of January 17, 2025, we identified $20,456$ out of $25,873$ Threads accounts ($79.1\%$) on \texttt{mastodon.social}, as verified by the instance administrator. We validated their presence in the Fediverse using WebFinger\footnote{\url{https://github.com/heliomass/Threads-Federation-Tracker}}, a decentralized protocol that retrieves user metadata from their domain. Our dataset includes $20,231$ Mastodon users from $1,417$ instances who interacted with these Threads users. To ensure privacy, user-specific attributes were encrypted at the field level.

\section{Novel Study on Cross-platform Interaction}
We denote $\mathcal{U}_T$ as the set of Threads users in our dataset who enabled Fediverse Sharing, and $\mathcal{U}_M$ as the set of Mastodon users who replied to posts made by $\mathcal{U}_T$.

\subsection{Characteristics of Users on Each Platform}

\begin{figure}[t]
  \centering
  \begin{minipage}[t]{0.48\textwidth}
    \centering
    \includegraphics[width=\textwidth]{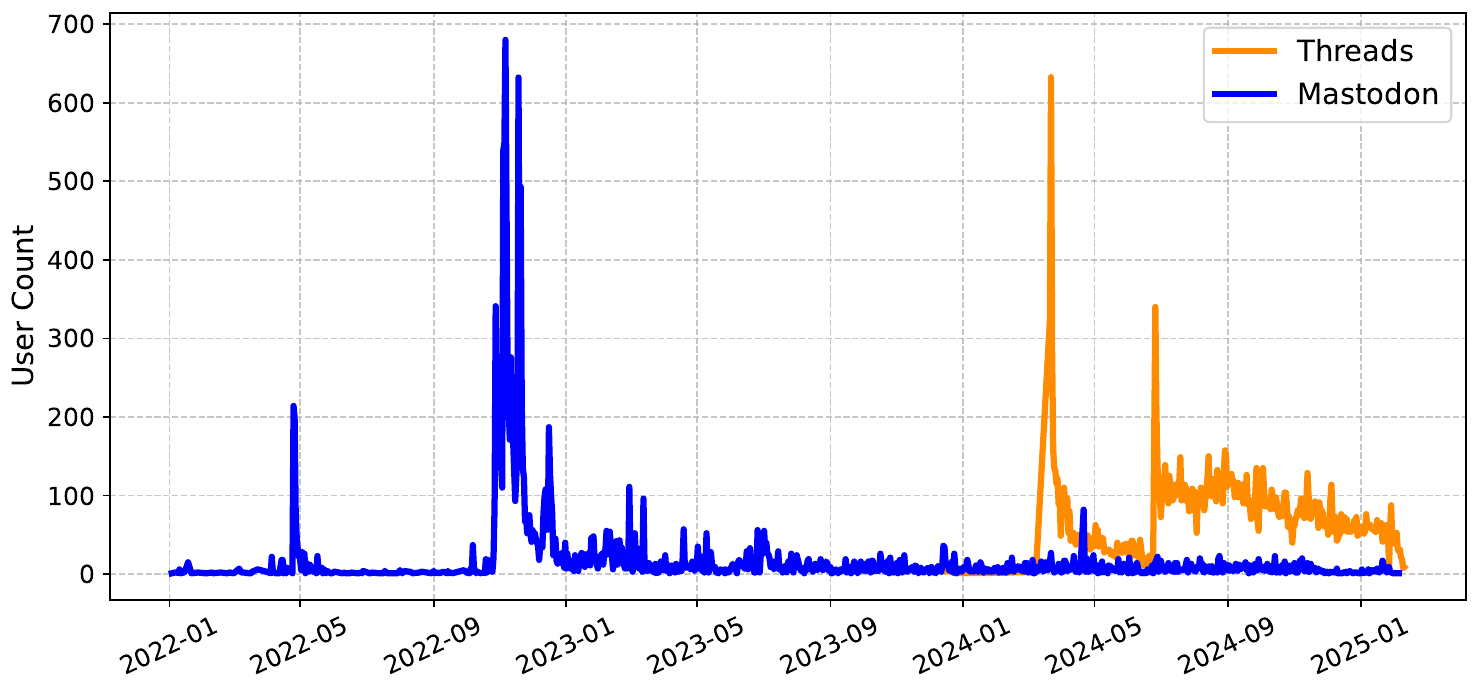}
    \caption{Joining dates of $\mathcal{U}_T$ and $\mathcal{U}_M$ to Mastodon: $\mathcal{U}_T$ join by enabling Fediverse Sharing, while $\mathcal{U}_M$ join by registering directly to a Mastodon instance.}
    \label{fig:fediverse_join_dates_by_threads}
  \end{minipage}%
  \hfill
  \begin{minipage}[t]{0.48\textwidth}
    \centering
    \includegraphics[width=\textwidth]{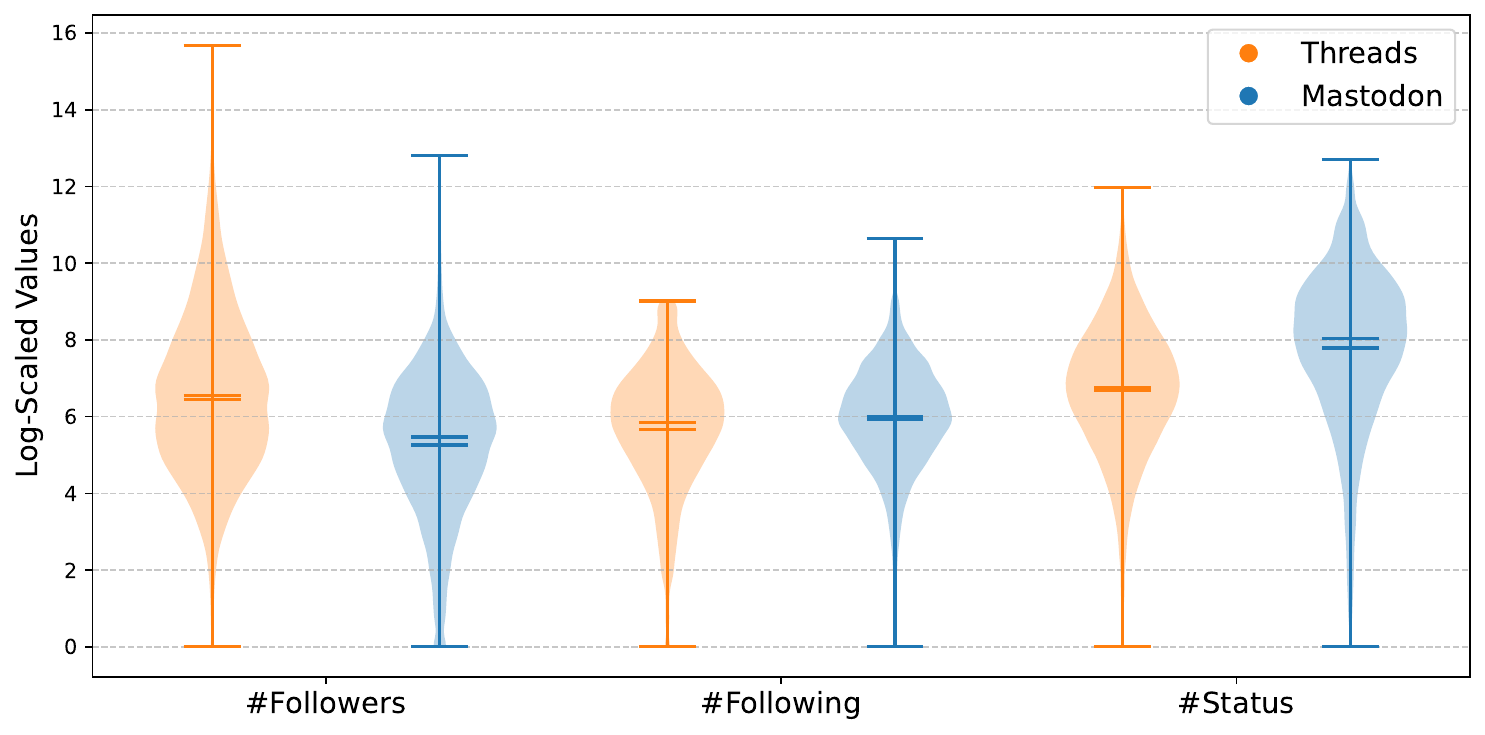}
    \caption{Violin plots comparing log-scaled profile metrics between $\mathcal{U}_M$ and $\mathcal{U}_T$. Each pair of distributions is significantly different by KS-test ($p<0.05$).}
    \label{fig:user_profile_metrics_compare}
  \end{minipage}
\end{figure}

\subsubsection{Joining Fediverse}

We analyzed account creation patterns to uncover the motivations driving Fediverse adoption. As shown in Figure~\ref{fig:fediverse_join_dates_by_threads}, Threads users show two notable surges: the first around the beta launch of Fediverse Sharing (March 17, 2024), and the second following the rollout of cross-platform interactions (July 5, 2024). In contrast, Mastodon sign-ups peaked on October 27, 2022, coinciding with Twitter’s leadership change. These patterns suggest that new platform features primarily drew Threads users, whereas Mastodon users were motivated by dissatisfaction with their previous platform~\cite{jeong2024exploring,cava2023drivers,he2023flocking}.

\subsubsection{Numerical Profile Attributes}

To compare user characteristics across platforms, we analyzed numerical attributes such as follower count, following count, and total statuses. Figure~\ref{fig:user_profile_metrics_compare} presents distinct behavioral patterns of  $\mathcal{U}_T$ and $\mathcal{U}_M$:

\begin{itemize}[topsep=2pt,itemsep=1pt,leftmargin=*]
\item \textbf{Followers:} Threads users exhibit a higher median and greater variability in follower counts, reflecting the platform's larger scale and stronger emphasis on visibility and follower accumulation.
\item \textbf{Following:} Mastodon users show a slightly higher median and broader distribution, suggesting more deliberate efforts to discover and connect with others in the absence of algorithmic recommendations.
\item \textbf{Status:} Mastodon users tend to post more frequently, with higher median and wider variation in total statuses, indicative of a culture that prioritizes engagement through activity and conversation.
\end{itemize}

These patterns reveal contrasting social dynamics: Threads users tend to focus on follower-oriented behavior, aligning with influencer culture, while Mastodon users typically engage in more proactive networking. These behaviors were similarly observed in prior studies~\cite{jeong2024exploring,jeong2024user,radivojevic2025user}.

\subsection{How Do Users Interact Across Platforms?}
\subsubsection{Daily Status Trend}
\begin{figure}
  \includegraphics[width=1.0\textwidth]{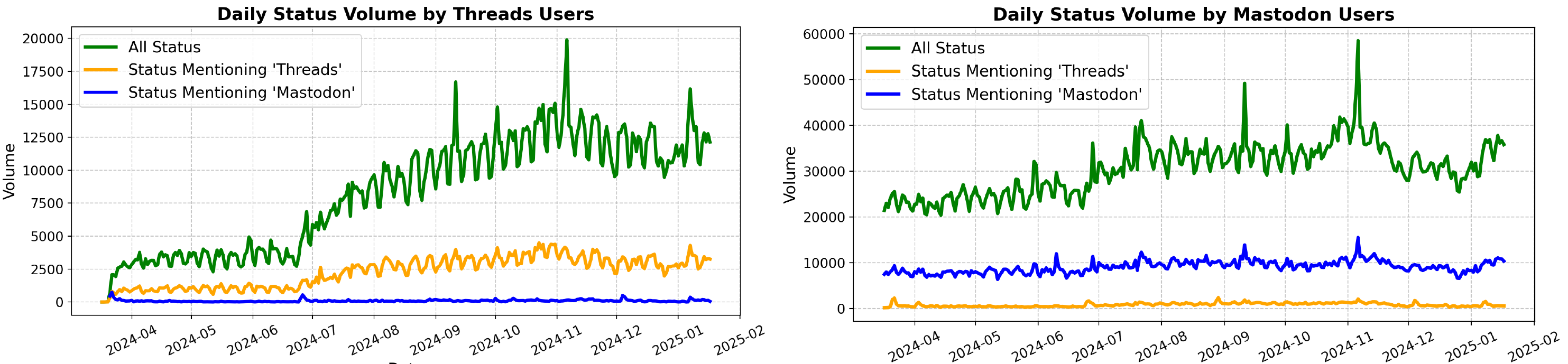}
\caption{Daily status trends for $\mathcal{U}_T$ and $\mathcal{U}_M$: green shows total statuses, orange marks those mentioning \texttt{threads.net}, and blue indicates those mentioning any of 1,417 Mastodon instances (e.g., \texttt{mastodon.social}, \texttt{mstdn.social}, etc).}
  \label{fig:metnion_trend}
\end{figure}

Figure~\ref{fig:metnion_trend} illustrates the daily volume of statuses from Threads and Mastodon users, along with statuses mentioning keywords related to the two platforms. After the beta release of Fediverse Sharing, we observed a gradual upward trend in status activity, which peaked following the second update on July 5th. This update introduced enhanced features such as users in Threads can like and reply to statuses in Mastodon.

To assess how users' attention towards platforms shifted over time, we analyzed the frequency with which they mentioned each platform in their statuses. Mentions of the other platform were consistently low, while users predominantly referred to their own. This self-referential pattern reflects identity-marking behavior, where individuals emphasize their group affiliation. Ethnogenesis research suggests that such explicit identity signaling is common during early cross-group interactions as a means of building trust~\cite{okamura1981situational}.

\subsubsection{Average Status Volume Shift}
\begin{table}[ht]
\centering
\small
\setlength{\tabcolsep}{4pt} 
\begin{tabular}{ccccccc}
\hline
\textbf{Time} & \textbf{Type} & \textbf{Avg. Before} & \textbf{Avg. After} & \textbf{Change} & \textbf{$t$-stat} & \textbf{$p$-val} \\
\hline
\multirow{2}{*}{$t_1$ (March 13th)} & Post  & 3.03 & 3.96 & 30.63\% & -11.11 & 0.000*** \\
                                & Reply & 1.57 & 1.80 & 14.45\% & -8.16  & 0.000*** \\
\hline
\multirow{2}{*}{$t_2$ (July 5th)}  & Post  & 3.17 & 4.35 & 37.22\% & -24.80 & 0.000*** \\
                                & Reply & 1.60 & 1.89 & 18.12\% & -15.33 & 0.000*** \\
\hline
\end{tabular}
\caption{Average status volume in $\mathcal{U}_M$ before and after Fediverse Sharing, measured at $t_1$ (beta-launch) and $t_2$ (post-update). $p$-value is rounded to three digits.}
\label{table:before_after_activity_level}
\end{table}

To account for potential shifts in the average number of statuses, we analyzed the status volume of $\mathcal{U}_M$ following the update in Fediverse Sharing. To enable meaningful comparisons, we first normalized the status volume by considering the number of users eligible to create a status. Here, the average activity levels before and after the updates were adjusted based on the duration of each period. The normalization was defined:

\[
\text{Normalized Status} = \frac{\text{\#Status}}{\text{\#Joined\:Users} \times \text{Days in Period}}
\]

Subsequently, we conducted a T-test on two distribution pairs. As shown in Table~\ref{table:before_after_activity_level}, the increase in post was more pronounced than in replies, likely due to the limited implementation of sequential replying. Nevertheless, both the average number of posts and replies demonstrated statistically significant increases ($p < 0.001$).  This analysis focused on the status of $\mathcal{U}_M$, as historical data for $\mathcal{U}_T$ is unavailable. Due to the API limit in Mastodon, the status for Threads users is only accessible from the point they joined federated networks, preventing retrospective analysis of their prior activity.

\subsubsection{Cross-platform Interaction Asymmetry}

\begin{figure}[t]
  \centering
  \begin{minipage}[t]{0.48\textwidth}
    \centering
    \includegraphics[width=\textwidth]{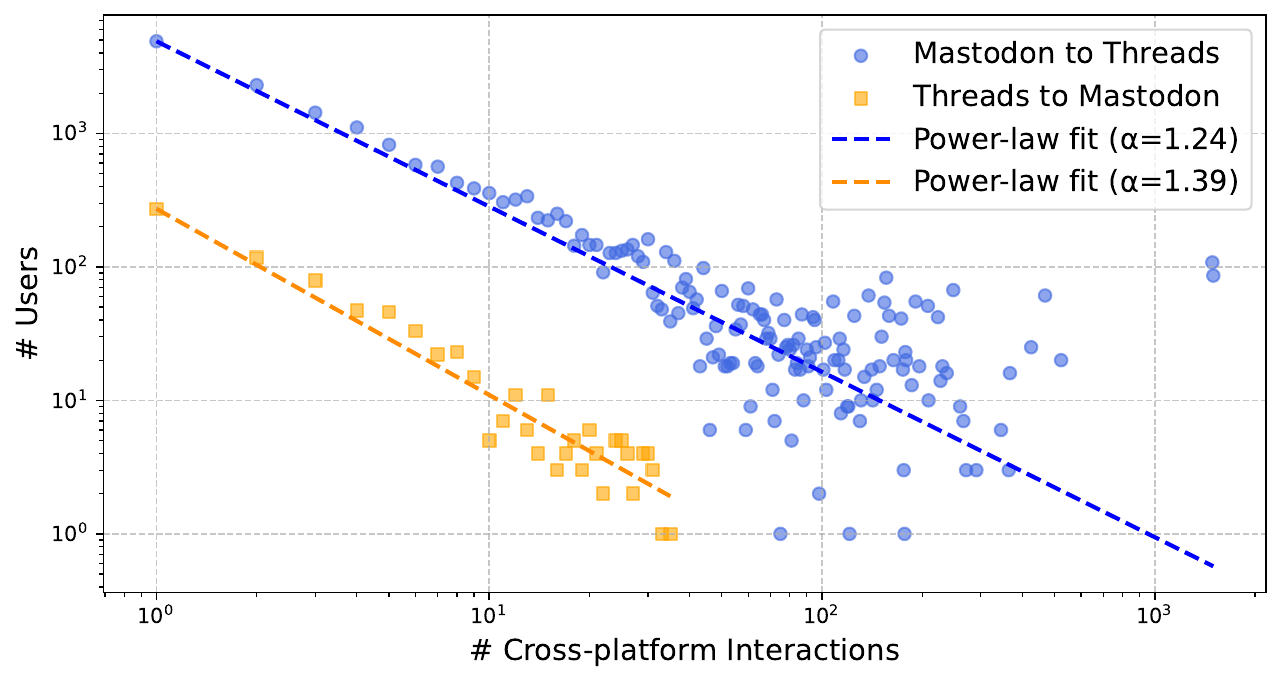}
    \caption{Scatter plot of cross-platform interactions between $\mathcal{U}_T$ and $\mathcal{U}_M$ via replies and likes. Both follow a power-law distribution.}
    \label{fig:cross_platform_interaction_scatter}
  \end{minipage}%
  \hfill
  \begin{minipage}[t]{0.48\textwidth}
    \centering
    \includegraphics[width=\textwidth]{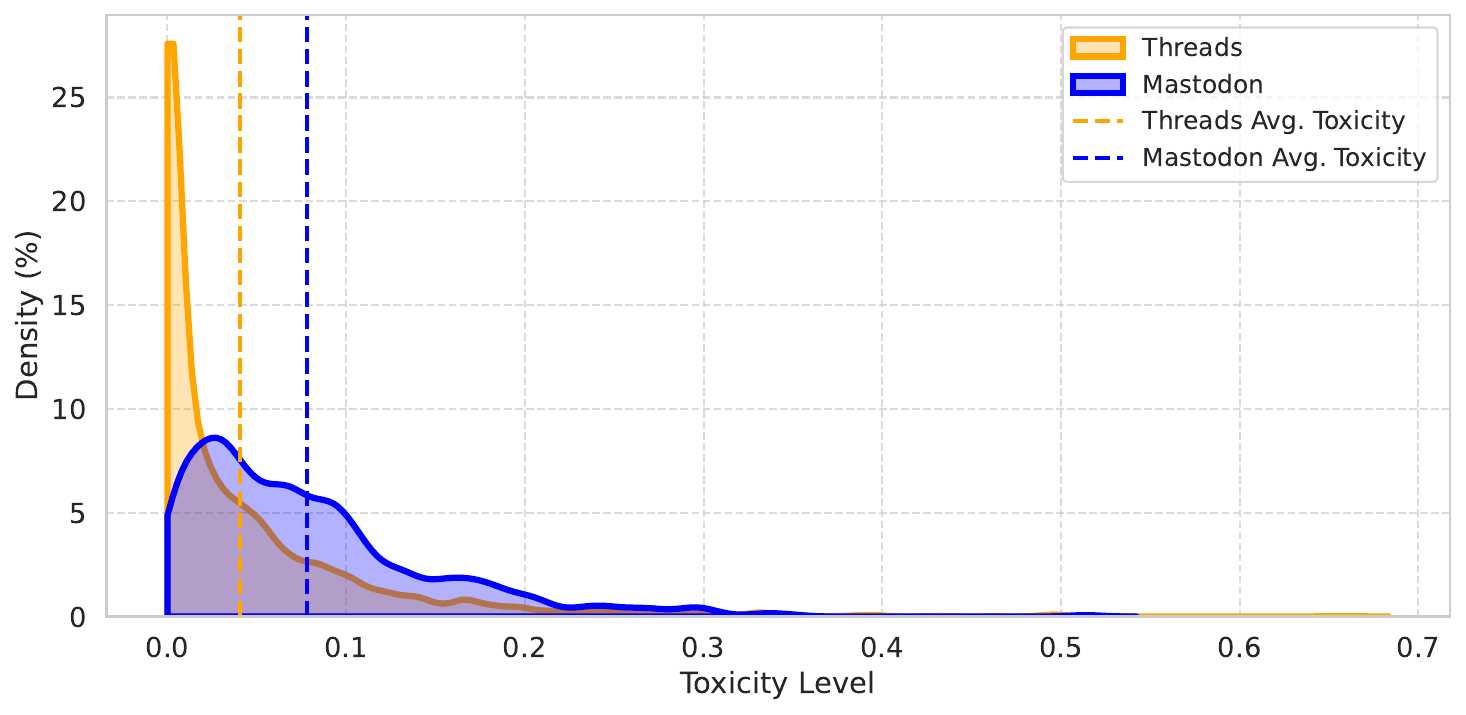}
    \caption{Average toxicity scores of users in $\mathcal{U}_T$ and $\mathcal{U}_M$, computed using their post history. Each user’s score is the mean toxicity across their posts.}
    \label{fig:toxicitycomparison_user_posts}
  \end{minipage}
\end{figure}

Figure~\ref{fig:cross_platform_interaction_scatter} shows that cross-platform interactions between Threads and Mastodon follow a power-law distribution, dominated by a few interactions. The estimated power-law exponent is $1.24$ for Mastodon-to-Threads and $1.39$ for Threads-to-Mastodon, highlighting distinct engagement patterns. Mastodon-to-Threads interactions are more frequent, while Threads-to-Mastodon interactions are significantly lower, indicating asymmetry. The smaller $\alpha$ for Mastodon-to-Threads indicates that cross-platform interaction is more concentrated among a few highly active users.





\subsection{What Do Users Express through Contents?}

\subsubsection{Toxicity across User Generated Contents}
Understanding how toxicity varies among users is important because levels of toxicity can reveal group-level identity, platform moderation style, and the overall quality of discourse. Encountering a broad spectrum of toxicity can present challenges for maintaining civil discussions, while also offering valuable insights into how various online communities will respond to different groups with hostility.

Figure~\ref{fig:toxicitycomparison_user_posts} presents the distribution of user-level toxicity scores, calculated using the Perspective API\footnote{\url{https://perspectiveapi.com}}. This multilingual model assigns a toxicity score from 0 to 1; we averaged these scores across each user's statuses to obtain their overall toxicity level. Our analysis reveals platform-level differences: while average toxicity remains low (below 0.1) on both platforms, Mastodon users exhibit a broader range and slightly higher levels. This likely stems from its decentralized structure, limited centralized moderation, and fewer content restrictions. In contrast, Threads’ centralized governance and corporate moderation contribute to a more consistently moderated, relatively less toxic environment.

\begin{figure}[t]
  \centering
  \begin{minipage}[t]{0.5\textwidth}
    \centering
    \includegraphics[width=\textwidth]{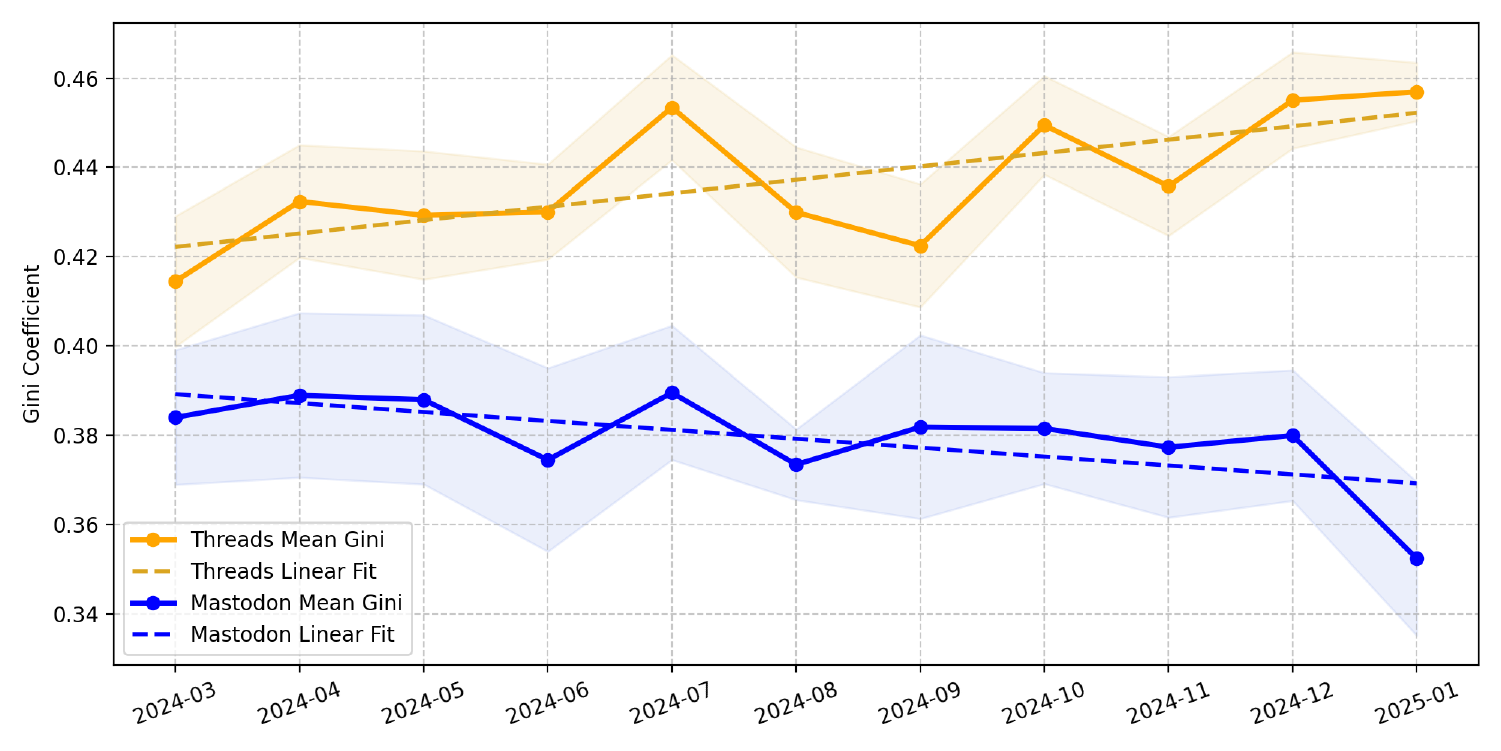}
    \caption{Gini coefficients (mean ± std) for topic distribution of status, comparing  $\mathcal{U}_T$ and $\mathcal{U}_M$. Linear regression shows the evolving direction of topic imbalance.}
    \label{fig:gini_topic_trend}
  \end{minipage}%
  \hfill
  \begin{minipage}[t]{0.48\textwidth}
    \centering
    \includegraphics[width=\textwidth]{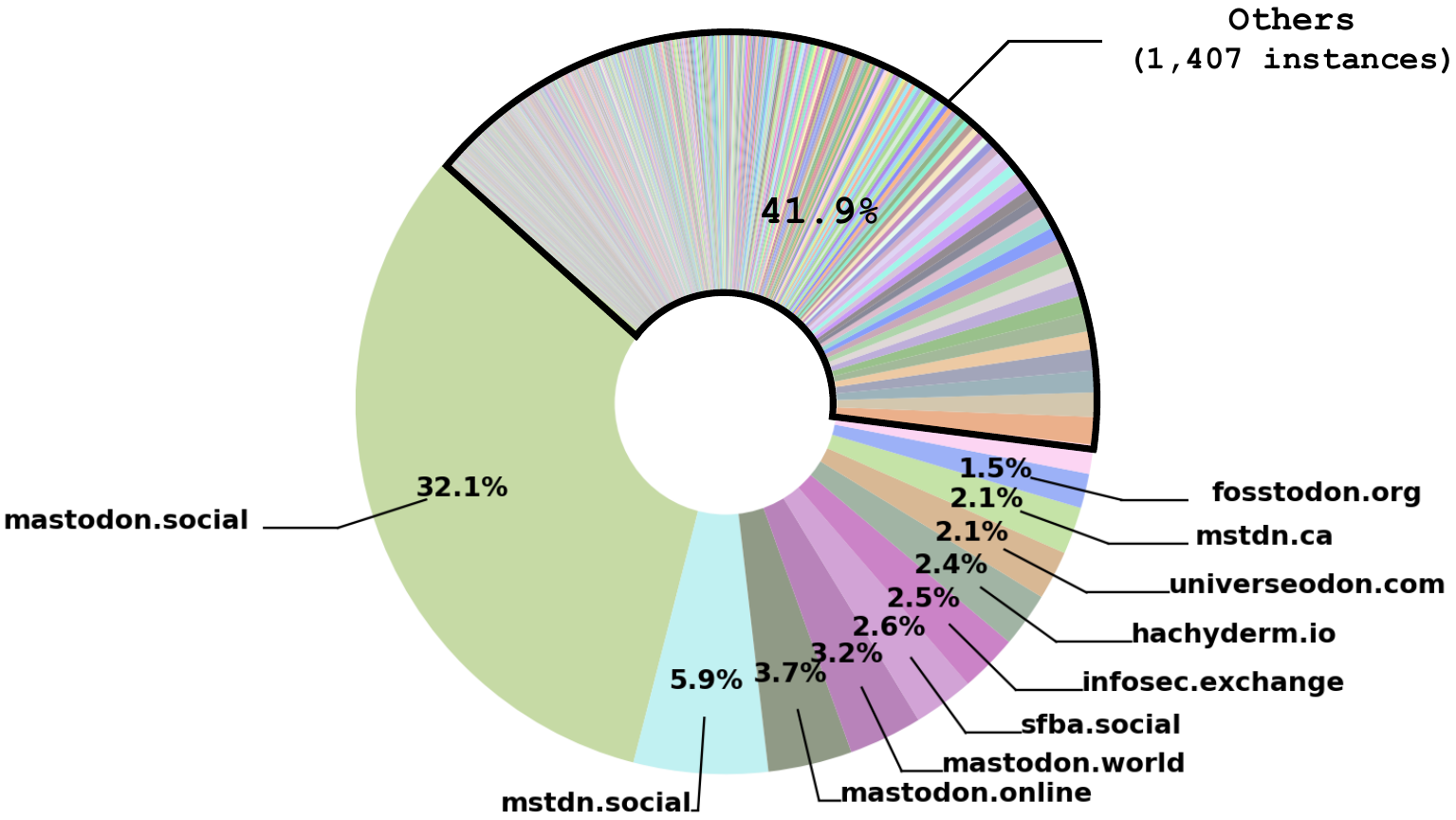}
    \caption{Distribution of home instances of interacted Mastodon users ($\mathcal{I}_M$). The top 10 instances are labeled with their user origin percentages.}
    \label{fig:mastodon_users_interacted_servers}
  \end{minipage}
\end{figure}

Figure~\ref{fig:gini_topic_trend} shows how topic diversity evolves over time for historical statuses from $\mathcal{U}_M$ and $\mathcal{U}_T$, measured using the Gini coefficient:

\[
Gini\:Coefficient = 1 - \sum_{i=1}^{n} p_i^2
\]
where \( p_i \) represents the proportion of status in cluster \( i \), and \( n \) is the total number of clusters. Since the number of statuses differs across platforms, we sampled 10K statuses per month to ensure balanced coverage. To capture variability from random sampling, this process was repeated ten times per month, and we report the mean and standard deviation of the resulting Gini coefficients.

Topic clustering was performed using BERTopic~\cite{grootendorst2022bertopic}, which combines sentence embeddings with a DBSCAN algorithm that does not require predefining the number of clusters. Details of the embedding model and clustering parameters are provided in the Appendix~\ref{appendix:topicmodeling}. After clustering, we excluded unassigned or sparse clusters (i.e., those with 10 or fewer samples) before computing the Gini coefficient to quantify monthly topic distribution.

As shown in Figure~\ref{fig:gini_topic_trend}, the Gini coefficient for Threads shows a slight upward trend in topic imbalance over time ($\beta_1 = 0.003$), while Mastodon shows a slight decline ($\beta_1 = -0.002$). This contrast calls for further investigation into how platform characteristics shape conversational dynamics after platform integration.

\subsubsection{Out-group Animosity Across Platforms}

To understand how users perceive the other platform, we analyzed posts mentioning either Mastodon or Threads. This is vital because out-group animosity can be related to increased polarization, creating barriers to conversation and hindering true platform integration.

We employed Aspect-Based Sentiment Analysis (ABSA) using a fine-tuned DeBERTa model~\cite{he2020deberta}, which categorizes sentiment into three classes (\textit{positive}, \textit{neutral}, or \textit{negative}) with respect to two targeted aspects (\textit{Threads} or \textit{Mastodon}). Further details on the model configuration are provided in the Appendix~\ref{appendix:sentimentanalysis}.

We formalized sentiment classification as \( \text{cls}(a, \mathcal{S}, \mathcal{T}) \), where \( \mathcal{S} \) is the source platform and \( \mathcal{T} \) the target platform. Our analysis was conducted on two levels:

\begin{enumerate}
    \item \textbf{Status-level}: For each status \( a \in \mathcal{A} \), we examined the distribution of sentiment labels \( \text{cls}(a, \mathcal{S}, \mathcal{T}) \) to determine how users express sentiment toward the target platform.
    
    \item \textbf{User-level}: For each user \( u \in \mathcal{U} \) with their corresponding set of statuses \( \mathcal{A}_u \), we computed the user’s overall sentiment toward the target platform based on the most frequent status labels.
    where the user's sentiment is determined by the most frequent sentiment label across their statuses.
\end{enumerate}

\begin{table}[ht]
\centering
\small
\setlength{\tabcolsep}{12pt} 
\begin{tabular}{ccccc}
\hline
\textbf{Source $\rightarrow$ Target} & \textbf{Level} & \textbf{Positive} & \textbf{Neutral} & \textbf{Negative} \\
\hline
\multirow{2}{*}{Mastodon $\rightarrow$ Threads} & Status & 7.66\% & 82.26\% & 10.07\% \\
                                                & User   & 8.93\% & 78.46\% & 12.61\% \\
\hline
\multirow{2}{*}{Threads $\rightarrow$ Mastodon} & Status & 22.12\% & 68.41\% & 9.47\% \\
                                                & User   & 27.36\% & 58.27\% & 14.38\% \\
\hline
\end{tabular}
\caption{Aspect-based sentiment distribution toward the two platforms, measured at both the individual status level and the user-aggregated level.}
\label{table:aspect_based_sentiment}
\end{table}

Table \ref{table:aspect_based_sentiment} illustrates the sentiment distribution between Threads and Mastodon users. We observed that Threads users expressed more positive sentiment towards Mastodon, while Mastodon users displayed a slightly more negative disposition towards Threads. Despite these differences, neutral sentiment dominated both platforms, presenting an opportunity for future shifts in perspective.

\subsection{What Instances Federate and Interact?}

Given a Mastodon user \( u \in \mathcal{U}_M \), each user is affiliated with one Mastodon instance \( i \). The set of every users' Mastodon instances, \( \mathcal{I}_M \), is denoted as:

\[
\mathcal{I}_M = \{ i \mid i = \text{home instance of } u, \; u \in \mathcal{U}_M \}
\]

\subsubsection{Instances Interacting with Threads}

Figure~\ref{fig:mastodon_users_interacted_servers} illustrates the distribution of home instances within $\mathcal{U}_M$. Among the $1,417$ Mastodon instances, a majority of users, specifically $63.1\%$ ($12,907$ out of $20,456$) of users in $\mathcal{U}_M$, are registered to \texttt{mastodon.social}, which is also recognized as one of the largest instances within the Mastodon network.

\subsubsection{Instance Size, Federation, and User Interaction}
We examined how instance size (measured by the number of registered users) relates to federation with Threads. Table~\ref{tab:coefficients} presents the statistical associations between instance size (as the dependent variable) and our three key independent variables.

\begin{itemize}
    \item \textbf{Cross-platform Interactions}: The total volume of interactions occurring between a Mastodon instance's users and Threads users.
    \item \textbf{Unique Interacted Users}: The count of users within an instance who interacted with Threads users at least once.
    \item \textbf{Federated Instances}: The number of other instances with which a given Mastodon instance has established federation.
\end{itemize}

\begin{table}[ht]
\centering
\small
\setlength{\tabcolsep}{12pt} 
\begin{tabular}{ccccc}
\hline
\textbf{Feature} & \textbf{$\beta$} & \textbf{SE} & \textbf{$p$-value} & \textbf{OR} \\
\hline
Cross-Platform Interaction  & -0.667 & 0.144 & 0.000*** & 0.513  \\
Unique Interacted Users     & 2.426  & 0.177 & 0.000*** & 11.314 \\
Federated Instances         & 1.162  & 0.086 & 0.000*** & 3.196  \\
\hline
\end{tabular}
\caption{OLS regression results examining the relationship between instance size and three instance-level features. Coefficient estimates ($\beta$), standard errors (SE), and odds ratios (OR) are shown. All values are rounded to three decimals.}
\label{tab:coefficients}
\end{table}

The linear regression analysis reveals key insights into the relationship between Mastodon instance size and interaction with Threads. There is a negative relationship between cross-platform interaction volume and instance size ($\beta = -0.667$), indicating that larger instances do not proportionally increase total interactions, so smaller instances can maintain high activity. Conversely, the number of unique interacted users shows a strong positive association with instance size ($\beta = 2.426$), meaning larger instances provide more opportunities for individual users to engage with Threads. In addition, the number of federated instances is also positively related to instance size ($\beta = 1.162$), showing that larger Mastodon instances connect with more other instances, likely because many users can help build connections to other instances.

\section{Limitations}
Threads users remain undiscoverable via the Mastodon API unless a Mastodon user explicitly accesses their profile or content. Inactive Threads users leave no persistent traces on any instance, limiting their overall visibility within the network. Moreover, replies from Threads users who have not enabled Fediverse Sharing are inaccessible to Mastodon users. Furthermore, the follow feature introduced on December 2, 2024, could not be analyzed, as Threads restricts access to its social graph, rendering this dimension of interaction beyond the scope of our study. Last, interpreting toxicity analysis can be subjective, and tools like the Perspective API, despite their multilingual capabilities, may still exhibit inaccuracies in certain languages~\cite{nogara2025toxic}.

\section{Future Work}
We will establish a controlled methodology using difference-in-differences (DID) analysis to derive generalized causal effects in cross-platform interactions. We'll construct a control group by surveying Mastodon users unaware of Threads, determining if increased status update activity results from platform integration, and revealing how interconnected ecosystems shape user engagement. We will also examine whether trust-building behavior drives self-referencing in status updates, assessing trust as a causal factor in how users present their platform affiliation, given that ethnogenesis studies suggest trust drives identity disclosure in initial interactions. Furthermore, we will investigate whether behavioral differences between Threads and Mastodon users stem from their structural distinctions (centralized and decentralized), as these likely reflect variations in users’ goals and engagement, potentially impacting cross-platform interaction extent.
\section{Conclusion}

We studied Fediverse Sharing, a service bridging Threads and Mastodon to enable cross-platform interaction. Our study involved collecting $79.1\%$ of Threads accounts available on \texttt{mastodon.social}. Our analysis uncovers user behavior, engagement patterns, and content dynamics before and after the platform integration. The key findings are summarized as follows:

\begin{itemize}
    \item \textbf{Adoption motivations diverge:} Mastodon users migrated from Twitter, fostering their community through proactive networking, while Threads users, drawn by new features, typically centered on follower growth.
    
    \item \textbf{Interaction remains asymmetric:} Though status activity increased on both platforms post-Fediverse Sharing launch, cross-platform interactions were notably imbalanced, with Mastodon users engaging Threads users more.
    
    \item \textbf{Content dynamics vary:} While toxicity levels were generally low across both, Mastodon showed greater variability in this regard. We also observed divergent trends in topic diversity among users on each platform. Although sentiment remained predominantly neutral, it is worth noting that Threads users expressed more positive views toward Mastodon.
    
    \item \textbf{Scaling constraints emerge:} While most interactions with Threads originated from \texttt{mastodon.social}, the volume of these interactions did not scale with the instance's size. Because federation operates at the user level,  ongoing integration relies more heavily on the continuous engagement of users.
\end{itemize}

Our study introduces a dataset and analytical framework for examining cross-platform interactions within federated environments. We hope this work paves the way for future research on federated social networks and deepens our understanding of platform interoperability in the evolving social media landscape.


\section{Ethical Statement}\
This study was reviewed and approved as exempt by the Institutional Review Board (IRB) at Arizona State University. We performed a secondary analysis using only publicly available data. We strictly adhered to the terms of service for both Threads\footnote{\url{https://help.instagram.com/769983657850450}} and Mastodon\footnote{\url{https://mastodon.social/about}}, which included communicating with the administrator of \texttt{mastodon.social}. To safeguard user privacy, we only analyzed publicly searchable posts and profiles, explicitly excluding any private or followers-only content. All user-specific features, including username and ID, were anonymized prior to analysis. Our data collection scripts are available on our GitHub repository: \url{https://github.com/ujeong1/Fediverse_Sharing_ASONAM25/}.



\bibliographystyle{plain}
\bibliography{reference}

\appendix

        
        
        



\section{Appendix: Model Configurations}

\subsection{Topic Modeling}
\label{appendix:topicmodeling}
We used the \texttt{paraphrase-multilingual-MiniLM-L12-v2} embedding model with BERTopic. Dimensionality reduction was performed with UMAP using parameters: \texttt{n\_neighbors=15}, \texttt{n\_components=5}, \texttt{min\_dist=0}, and \texttt{metric=cosine}. Clustering was done via HDBSCAN with \texttt{min\_cluster\_size=30}, \texttt{min\_samples=10}, and \texttt{metric=euclidean}. Text vectorization employed \texttt{CountVectorizer} configured for bigrams (\texttt{ngram\_range=(1,2)}) and minimum document frequency 10 (\texttt{min\_df=10}). \texttt{reduce\_topics} was applied to merge semantically similar clusters.

\subsection{Aspect-Based Sentiment Analysis}
\label{appendix:sentimentanalysis}
We utilized the \texttt{deberta-v3-large-absa} model fine-tuned for target-oriented sentiment classification on two aspects: Threads and Mastodon. The model has a hidden size of 768, with 12 transformer layers and 12 attention heads.

\end{document}